\documentclass[12pt,preprint]{aastex}

\shorttitle{Electron-Positron Outflow from an AGN Fireball}

\shortauthors{Asano \& Takahara}

\def\siml{\lower4pt \hbox{$\buildrel < \over \sim$}}
\def\simg{\lower4pt \hbox{$\buildrel > \over \sim$}}

\begin{document}

\title{
A Relativistic Electron-Positron Outflow from a Tepid Fireball
}

\author{\scshape Katsuaki Asano\altaffilmark{1} and Fumio 
Takahara\altaffilmark{2}}
\email{asano@phys.titech.ac.jp, takahara@vega.ess.sci.osaka-u.ac.jp}


\altaffiltext{1}{Interactive Research Center of Science, 
Graduate School of Science,
Tokyo Institute of Technology,
2-12-1 Ookayama, Meguro-ku, Tokyo 152-8550, Japan}
\altaffiltext{2}{Department of Earth and Space Science, 
Graduate School of Science,
Osaka University,
Toyonaka 560-0043, Japan}

\begin{abstract}

Through detailed numerical simulations,
we demonstrate that relativistic outflows (Lorentz factor
$\Gamma \sim 7$) of electron-positron pairs
can be produced by radiative acceleration 
even when the flow starts from a nearly pair equilibrium state at 
subrelativistic temperatures.
Contrary to the expectation that pairs annihilate during 
an expansion stage for such low temperatures, we find that 
most pairs can survive for 
the situations obtained in our previous work.
This is because in the outflow-generating region 
the dynamical timescale is short enough even though the fireball is optically 
thick to scattering. 
Several problems that should be solved to apply to actual active galactic nucleus jets
are discussed.

\end{abstract}


\maketitle

\section{Introduction}
\label{sec:intro}

One of the most challenging problems in astrophysics is
the acceleration mechanism of relativistic jets from active 
galactic nuclei (AGNs)
and Galactic black hole candidates.
The bulk Lorentz factor of these jets is above 10 and 
the kinetic power is almost comparable to the
Eddington luminosity.
Although various jet models have been proposed,
we do not yet have a satisfactory solution.
Recent remarkable progress in magnetohydrodynamical (MHD)
simulations includes Poynting-dominated jet formation by
a rapidly rotating black hole with an accretion 
disk \citep[e.g.][]{mck06,haw06}.
Along the spin axis a centrifugal funnel filled with magnetic field
is formed.
The magnetic field in the funnel is considered to be amplified by
the differential rotation of the accretion disk
and the frame-dragging effect of the black hole.
However, in the simulation of \citet{mck06},
the total jet luminosity within the funnel is only 0.2\%
of the rest-mass energy accretion rate.
While continuous developments in
numerical studies are expected to probe MHD acceleration,
other types of acceleration mechanisms are worth considering.

One of the alternative acceleration mechanisms is the thermal expansion of
optically thick fireballs.
If the matter content of jets is electron-positron pairs,
thermal radiation pressure from accretion disks
may be sufficient to accelerate the jets.
The original model of the fireball \citep{ree92}
relevant to gamma-ray bursts (GRBs)
assumes that its initial stage is
radiation-dominated plasmas that are in complete thermal equilibrium
at high temperatures comparable to the electron rest-mass energy
$m_{\rm e} c^2$.
However, the characteristic size of AGNs is too large to achieve
complete thermal equilibrium.
The novel model proposed to overcome this problem
is electron-positron pair outflow from 
a ``Wien fireball'' \citep{iwa02,iwa04},
which is a photon-dominated, optically thick plasma,
but the densities of photons and pairs are less than those
in the complete thermal equilibrium fireballs.
Pairs coupled with photons by scattering
are thermally accelerated,
expending the internal energy of the fireball.
As is seen in the original model of the fireball,
the Lorentz factor $\Gamma=1/\sqrt{1-\beta^2}$ increases with the radius $R$,
and the temperature drops as $T \propto R^{-1}$.
\citet{iwa02,iwa04} showed that
a sufficient amount of electron-positron pairs
survive as a relativistic outflow, if
the temperature of the fireball at the photosphere is relativistic
($\gtrsim m_{\rm e} c^2$).

When the temperature is relativistic, all the cross-sections
of pair creation, annihilation and Compton scattering
are of the same order.
Therefore, a balance between pair creation and annihilation is realized,
and the number densities of pairs and photons are of the same order
as long as photons and pairs are coupled with each other.
On the other hand, the condition to make the pair-annihilation 
timescale shorter than
the dynamical timescale, 
\begin{equation}
n_+ \sigma_{\rm T} \frac{R}{\Gamma \beta}>
\frac{3}{8} \left[ 1+\frac{2 \theta^2}{\ln{(1.12\theta+1.3)}}\right],
\end{equation}
is almost the same as the condition of being optically thick to scattering,
where $n_+$, $\sigma_{\rm T}$,
and $\theta=T/m_{\rm e} c^2$ are the positron density
in the comoving frame, Thomson cross section,
and the normalized temperature, respectively.
This implies that
the number of pairs is almost conserved outside the photosphere, where
photons and pairs are decoupled.
Therefore, a relativistic temperature at the photosphere is required
to obtain a sufficient amount of pairs avoiding annihilation.

The temperature of fireballs is determined by microscopic processes
in their formation sites, presumably accretion disks,
such as radiative cooling, Compton scattering,
and $\gamma-\gamma$ pair production \citep[see e.g.][]{sve84}.
Assuming that pairs are confined with protons,
several authors \citep[see e.g.][and references there in]{kus87,bjo92}
investigated pair equilibrium plasmas.
However, a series of studies of pair equilibrium plasmas
implies that the equilibrium temperature is too low 
for the Wien fireball model.
The plasma temperature typically becomes $\theta \sim 0.1$
for plasmas of size $\sim 10^{14}$ cm
and luminosity $\gtrsim 10^{45} {\rm erg}$ ${\rm s}^{-1}$,
while the Wien fireball model
requires a super-Eddington luminosity
$\gtrsim 10^{47}$ erg ${\rm s^{-1}}$
(under the assumption of spherical symmetry)
and a high temperature $\theta > 0.5$.

The previous studies on pair equilibrium plasmas have not taken into account
the dynamical effects of outflowing electron-positron pairs
on the plasma temperature.
It is not trivial to estimate the motional (or advective)
effects of fireballs on radiative transfer, cooling processes,
pair density and temperature. In our previous paper 
\citep[][hereafter AT07]{asa07},
we simultaneously solved the dynamics of outflowing pairs, microphysics,
and radiative transfer to obtain the internal structure of fireballs.
Even in this simulation, we obtained only $\theta \sim 0.2$ at $R \sim$
several times $10^{14}$ cm
for the luminosity $L = 10^{47}$ erg ${\rm s^{-1}}$,
though a considerable amount of pairs outflow
with a mildly relativistic velocity.
In the simulations of AT07,
the acceleration of outflows is not finished within the simulation scale.
We pessimistically supposed that a rapid extinction of pairs inside the photosphere may
occur, since the obtained temperature is so low.

However, the detailed situations in AT07 are different from the simplified
inner boundary conditions of the outflow in the simulations of \citet{iwa04}.
The pair-photon ratio at the outer boundary of the simulation region 
is smaller than that of the Wien equilibrium plasma
by a factor of $\sim 2$.
The photon spectrum obtained numerically is also different from
the simple Wien spectrum.
It may be valuable to simulate subsequent evolution of the outflows
outside  such a tepid fireball obtained in the simulations in AT07.

In this Letter, we demonstrate that the tepid fireball obtained by
the simulations of AT07 can be accelerated to a relativistic velocity
conserving the pair number, even though the fireball temperature
is not high enough.
In \S \ref{sec:method}, we describe our simulation method, and 
show our results in
\S \ref{sec:results}.
\S \ref{sec:disc} is devoted to discussion.

\section{Method}

\label{sec:method}

\subsection{Energy-Release Region}

We numerically obtain spherically symmetric,
steady solutions of radiation and pair outflows.
As the inner boundary condition, we employ the numerical values 
obtained at the outer boundary in the simulations of AT07.
Namely, the simulations of AT07 provide us with the formation stage of the fireball.
First of all, let us review the physical situations in AT07.
In AT07, the proton number density is assumed to be
\begin{equation}
n_{\it p}(R) =
n_0 \exp\left[-(R/R_0)^2 \right],
\end{equation}
where $R_0$ and $n_0$ are constant.
In this region the plasma was assumed to be heated,
which mimics energy release via viscous heating.
The heating rate of the plasma was proportional to $n_{\it p}$.
The plasmas were divided into three fluids:
proton ({\it p}), background electron ({\it e}), and 
pair ({\it e$^\pm$}) fluids.
The background electrons and protons were assumed
to be static; the gravitational force of the central black hole
may arrest the background plasma.
Although the validity of this multifluid approximation is not always assured,
we had adopted this approximation for simplicity.
For the microphysics, Coulomb scattering, Compton scattering,
bremsstrahlung, and electron-positron
pair annihilation and creation were taken into account.
The frictional forces and heating of the pair plasma due to Coulomb scattering
with background fluids were calculated using numerical results
from \citet{asa07b}.
The effect of the frictional force was less important than
the radiative force.
The spherical symmetry in the geometry and the mildly relativistic outflow
yield beamed photon distribution.
It is a well-known effect that too fast flows are decelerated even by
a beamed radiation field (Compton drag).
The flow velocity was self-regulated by the beamed photon field.
Radiative transfer was solved with the Monte Carlo method.
Taking into account Compton scattering
and pair creation, the evolutions of energy and direction
along their trajectories are fully solved for each photon.
The energy range of photons $x \equiv \varepsilon/m_{\rm e} c^2$
was from $10^{-5}$ to $10^2$, and
the outer boundary was set at $R=R_{\rm out} \equiv 2 R_0$.
For the total heating rate $L=10^{47}$ erg ${\rm s^{-1}}$, AT07 obtained
mildly relativistic ($U \equiv \Gamma \beta \sim 1$) outflows.
The ratios of the total luminosity to the mass ejection rate,
$\eta \equiv L/4 \pi R_{\rm out}^2 n_\pm(R_{\rm out})
U(R_{\rm out}) m_{\rm e} c^3$, are $\sim 30$,
but the temperatures are relatively low ($\theta \sim 0.2$).

\subsection{Outer Region}

Using $U$, $n_\pm$, $\theta$ and beamed photon field at the outer
boundary in AT07 as the inner boundary conditions, 
we simulate the behavior outside the energy-release
region (the simulation region of AT07).
The detailed method of the simulations is the same as that in AT07.
However, we assume that protons do not exist anymore,
so we switch off heating and the Coulomb friction due
to background protons and electrons.
Thus, we take into account only Compton scattering,
bremsstrahlung, and electron-positron pair annihilation and creation
for the outer region.
We solve hydrodynamics and radiative transfer only for pair fluids.

\section{Results}

\label{sec:results}

\begin{figure}[t]
\centering
\epsscale{1.0}
\plotone{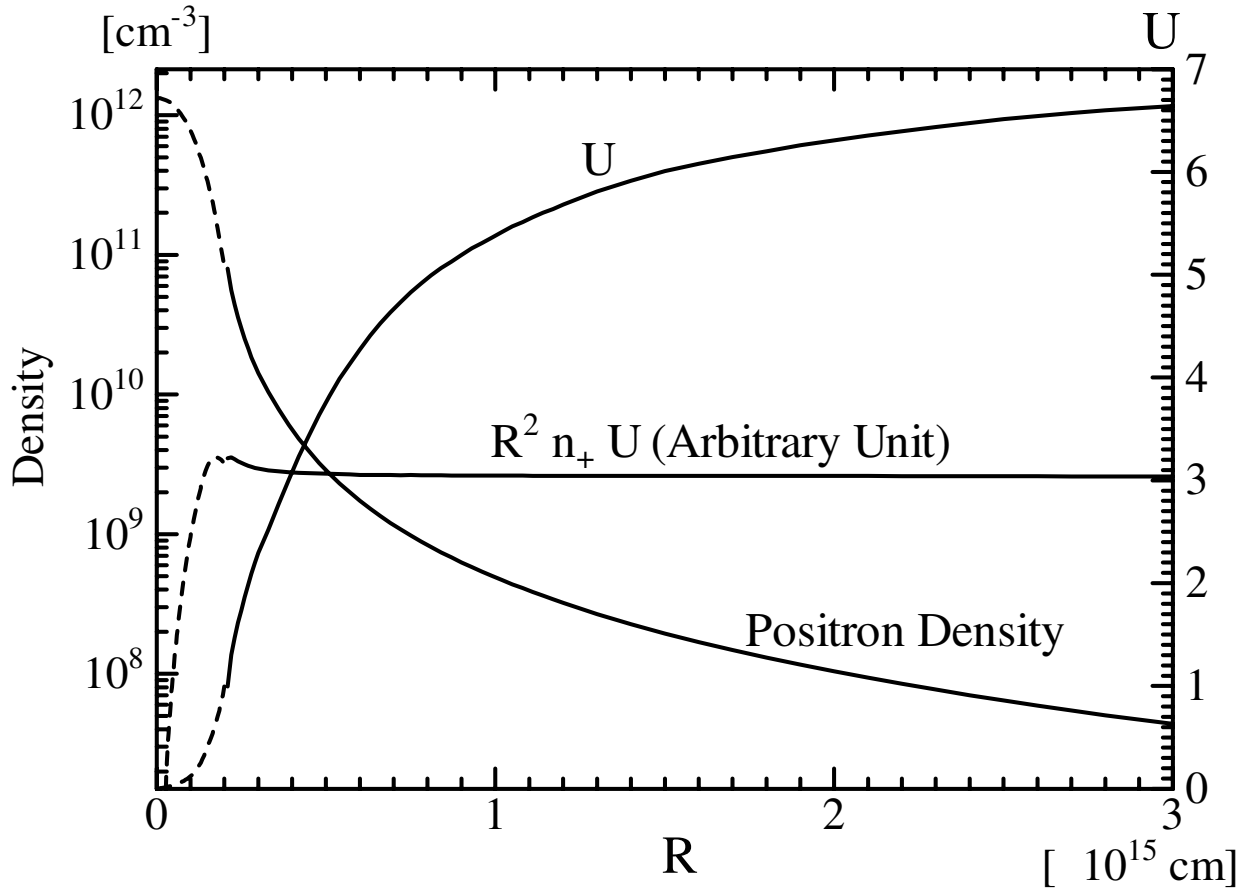}
\caption{Distributions of the positron density in the comoving frame 
$n_+$ (left axis),
4-velocity $U$ (right axis), and $R^2 n_+ U$ (arbitrary units,
see the left axis for reference). The dashed lines are the results from AT07.
\label{fig:density}}
\end{figure}

\begin{figure}[t]
\centering
\epsscale{1.0}
\plotone{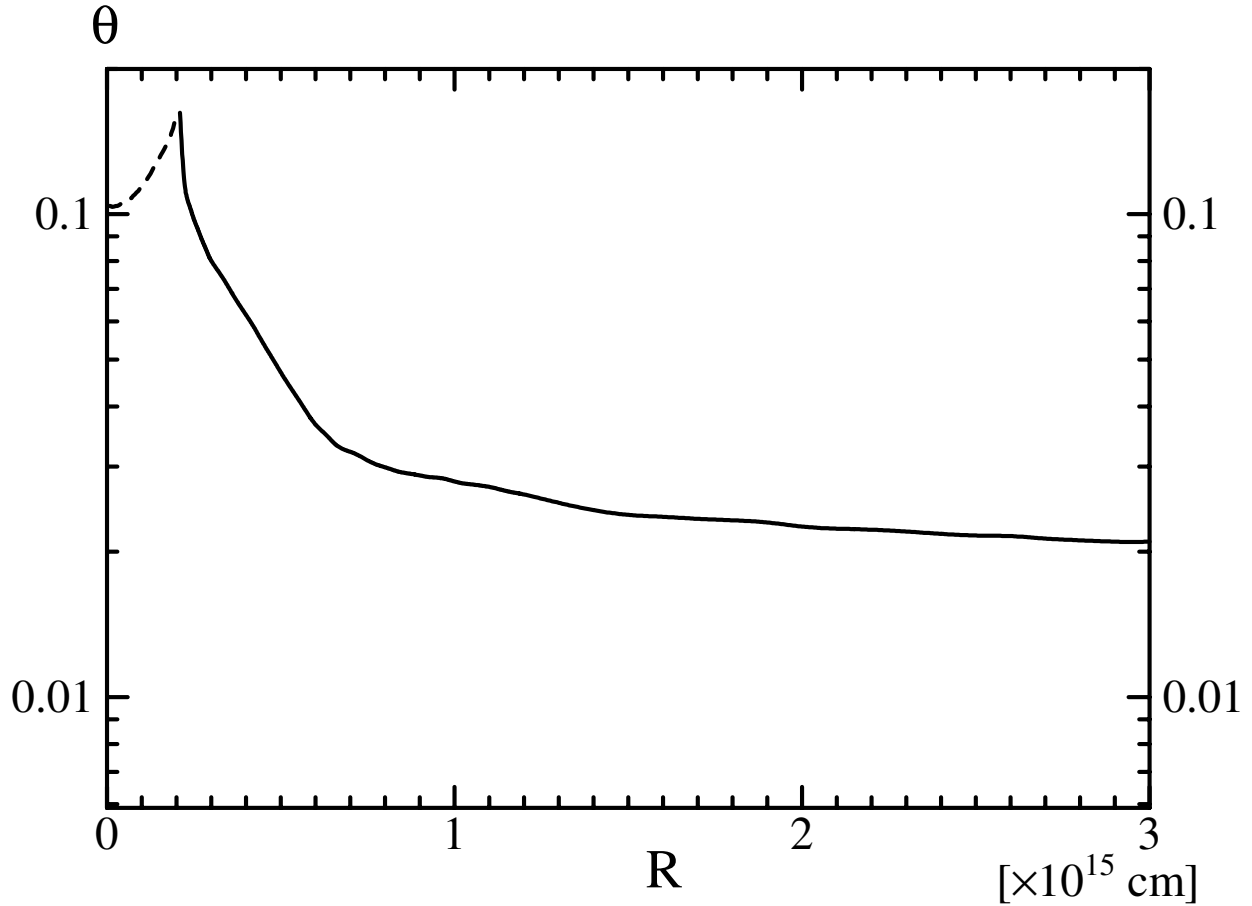}
\caption{Temperature of the pair plasma. The dashed line is 
the result from AT07.
\label{fig:temp}}
\end{figure}

In AT07 two sets of parameters were adopted: $R_0=10^{14}$ cm with
$n_0=10^{10}$ ${\rm cm^{-3}}$,
and $R_0=3 \times 10^{14}$ cm with $n_0=\frac{1}{3} \times 10^{10}$
${\rm cm^{-3}}$.
These values are characteristic of accretion plasmas in AGNs.
In both cases, the Thomson scattering depth due to the {\it e}-fluid
$\tau_{\rm p} \sim n_0 R_0 \sigma_{\rm T} \sim 0.7$.
The basic features of the obtained outflows are similar in these two cases.
The resultant density of the pair plasma overwhelms the density of the
background plasma so that $n_0$ may not be a critical parameter.
Although the results for the latter parameter set were explained
in detail in AT07, we show the results employing the former compact
case as an energy-release region in Figs. \ref{fig:density} and \ref{fig:temp}. 
The outflow is accelerated mainly by a radiative force.
While the photosphere is located at $\sim 10^{15}$ cm,
the acceleration turns out to continue even outside the photosphere.
Even though the fraction of photons interacting with the plasma
is small, the influence of radiative interaction outside the photosphere
cannot be neglected because of the copious amount of photons.
The final Lorentz factor becomes $\sim 7$.

Fig. \ref{fig:temp} represents the behavior of the electron temperature
and exhibits our rather artificial inner boundary conditions in 
the simulation.
Within $R=2 R_0$ (the boundary between the energy-release region
and the outer region),
the heating from the background protons makes electron temperature 
increase with radial distance.
The sudden shutdown of the heating at $R=2 R_0$ results in
the discrete change of temperature gradient.
Outside the energy-release region the temperature decreases monotonically,
but the temperature drop begins to slow down around the photosphere.
Outside the photosphere, the simple analytical formulae for fireballs
($U \propto R$ and $T \propto R^{-1}$) are no longer applicable.

If the plasma is in the Wien equilibrium,
a drastic extinction of pairs should occur for such a low temperature.
We also plot the positron number flux $R^2 n_+ U$ in Fig. \ref{fig:density}.
The dashed line (result from AT07) shows that most of the outflowing material
is provided in the outer part of the energy-release region.
Outside the energy-release region, even though the temperature decreases
from $\theta \sim 0.2$ to $0.03$ at the photosphere,
the number of pairs is almost conserved.
At $R=2 R_0$, the pair-annihilation timescale
is estimated as $\sim 2 \times 10^3$ s in the comoving frame,
while the timescale of density drop ($n_+$ becomes half in this timescale)
due to plasma expansion is $\sim 10^3$ s.
Thus, the two timescales are already comparable even though
$2 R_0$ is inside the photosphere.
Since the pair-annihilation timescale $\propto n_+^{-1}$,
we can neglect the pair annihilation outside $2 R_0$.
Photons and pairs are not in the Wien equilibrium,
though the plasma is optically thick to scattering.
This is because the pair plasma keeps on being heated during the initial acceleration,
and is smoothly evacuated outward at a mildly relativistic velocity.
If the temperature drop begins further inside, lower electron temperature
may lead to rapid annihilation of pairs.

\begin{figure}[h]
\centering
\epsscale{1.0}
\plotone{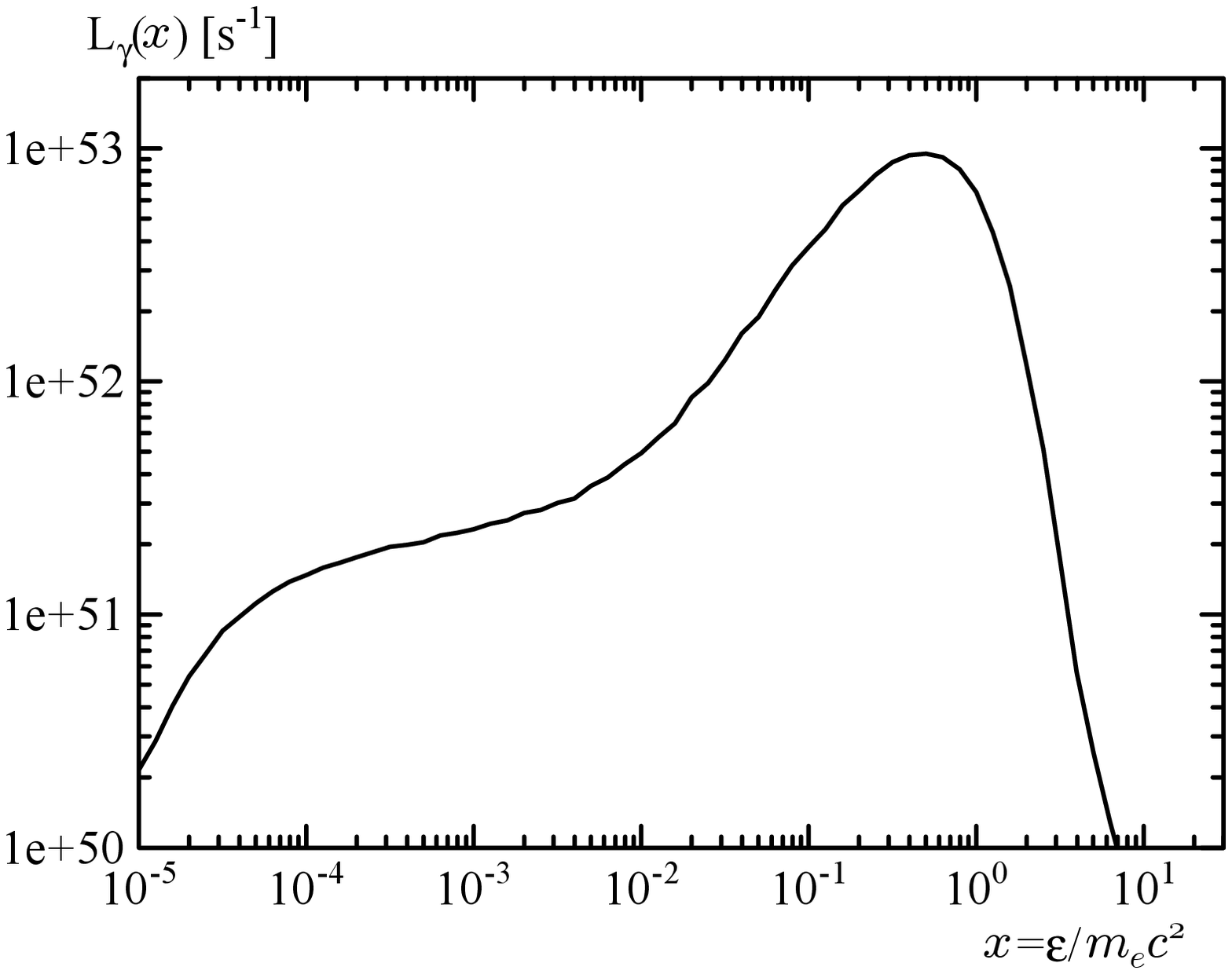}
\caption{Energy spectrum of the photons escaping from the outer boundary.
\label{fig:spec}}
\end{figure}

\begin{figure}[h]
\centering
\epsscale{1.0}
\plotone{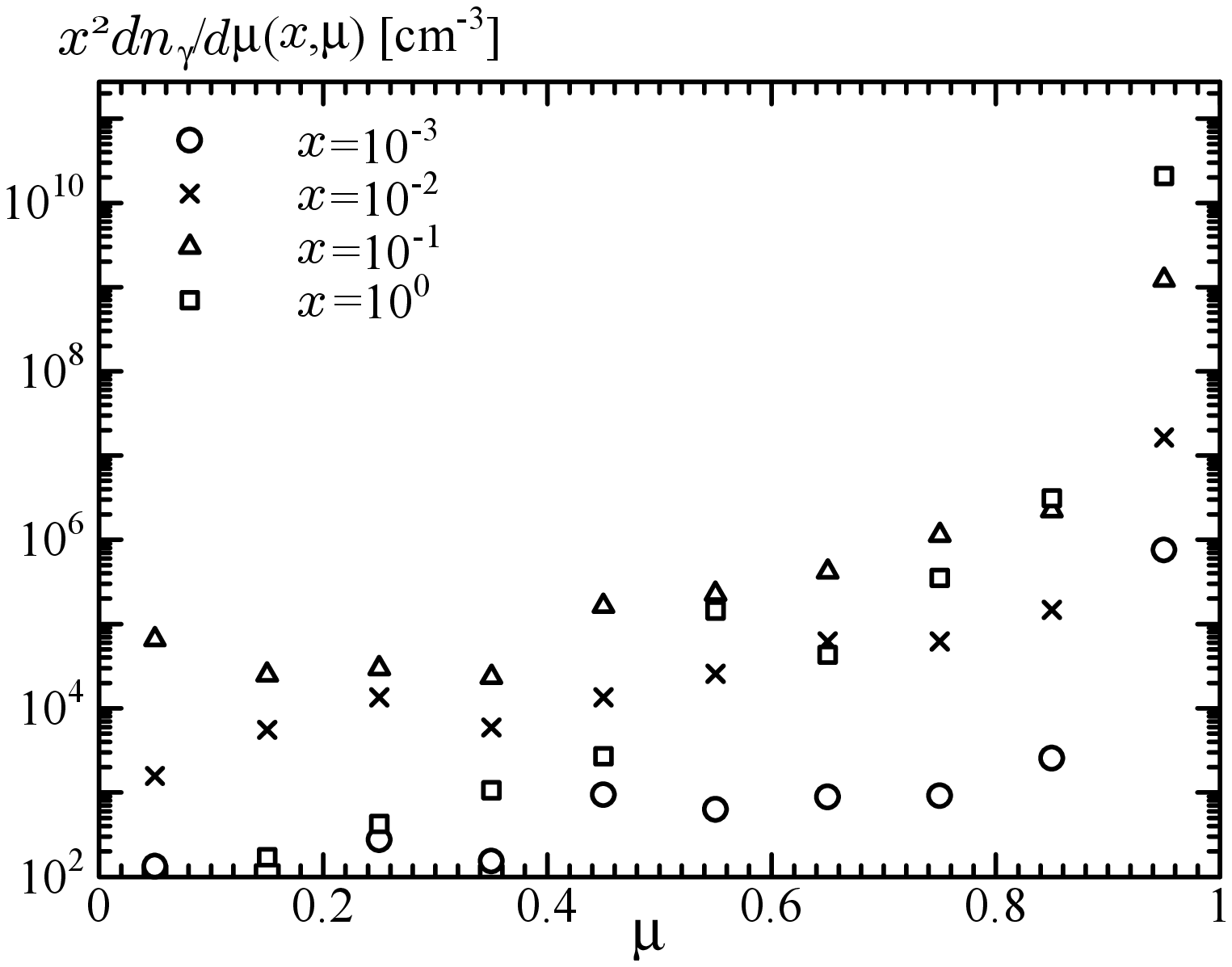}
\caption{Angular distribution of photons at the outer boundary,
where $\mu$ is the cosine between the direction of the photon
and the radial direction.
The results are plotted from binned data with $\Delta \mu=0.1$
after following $1.2 \times 10^9$ paths of photons.
\label{fig:ang}}
\end{figure}

A problem is that the energy outflow rate of the pair plasma
is only $\sim 1/60$ of the total luminosity $L$.
The rest of the energy escapes as photons from the boundary.
In Fig. \ref{fig:spec}, we plot the energy spectrum of the photons
escaping from the outer boundary of this simulation $L_\gamma(x) dx$
($x \equiv \varepsilon/m_{\rm e} c^2$).
The spectrum has a broader feature than that in the Wien spectrum.
There are huge amounts of soft photons ($x<10^{-2}$)
in comparison with the thermal spectrum.
As shown in Fig. \ref{fig:ang}, the photon field is highly beamed.
As for X-ray photons ($x \sim 10^{-2}$), the off-axis
flux is diluted by a factor of $\sim 100$ compared with
the on-axis flux.

We also simulate an outflow for the broader profile of protons:
$R_0=3 \times 10^{14}$ cm with $n_0=\frac{1}{3} \times 10^{10}$
${\rm cm^{-3}}$.
The qualitative results are the same as the results described above,
but the final Lorentz factor $\Gamma \sim 5$.
Given the luminosity $L$, the radius of the photosphere may not change
drastically.
Therefore, considering the behavior $\Gamma \propto R$,
the larger initial radius (namely the radius at $U \sim 1$)
results in a smaller Lorentz factor.
On the other hand, the energy outflow rate of the pair plasma
is 14 \% of the total luminosity $L$, which is larger than
the former case.
While a more compact energy release is preferable to increase
the final Lorentz factor, conversely the efficiency of the energy conversion
into pair outflow becomes larger for a more capacious source.

\section{Discussion}
\label{sec:disc}

Our simulation shows that an energy release of $10^{47}$ erg ${\rm s^{-1}}$
within a size of $\sim 10^{14}$ cm produces a pair outflow of $\Gamma \sim 7$.
Even though the fireball temperature is tepid,
rapid annihilation of pairs is avoided.
If the entire system we have considered is moving with a Lorentz factor
$\Gamma_{\rm s} \geq 1.06$ ($\beta_{\rm s} \geq 0.34$),
the Lorentz factor of the pair plasma seems to be larger than 10.
Such a nonrelativistic outflow of protons may be easily achieved,
as various MHD simulations show.
Therefore, similar simulations to ours
including the outflow of protons are valuable, while we have assumed
that the background protons are static.

The reason why a rapid annihilation of pairs does not occur may be
a slight difference in pair density from the prediction
of the Wien equilibrium.
This difference comes from the finite size of the energy-release region.
In the energy-release region the outflowing material keeps on being
heated and is evacuated with a mildly relativistic velocity.
The rapid expansion makes the pair plasma escape smoothly from
the optically thick region, avoiding pair annihilation.
For such a marginal optical depth, the Wien equilibrium is no longer
a reasonable approximation.
The pair densities obtained numerically
are suitable for avoiding pair annihilation
and keeping the plasma optically thick to scattering, 
which are opposing requirements 
in general.

However, resultant densities are too low to convert
the radiation energy into the kinetic energy of the pair plasma efficiently.
The efficiency in our simulation is only 1/60 of the total luminosity.
The results may depend on the profiles of the plasma heating rate.
Simulations for other types of profiles of the plasma heating rate
are worth studying.
We do not simulate for $L>10^{47}$ erg ${\rm s^{-1}}$,
because the higher pair density requires high computational cost,
while the temperature tends to decrease with $L$.
This is another challenge for us, though there is a possibility
of rapid extinction of pairs due to the lower temperature.

Of course, the most idealized method to efficiently produce relativistic 
outflows
is to create fireballs with relativistic temperatures.
Relativistic temperatures for optically thick plasmas assure
a higher density of pairs, which is preferable for
both the energy efficiency and the final Lorentz factor.
However, as AT07 showed for steady solutions,
even if the effects of mildly relativistic motion
are taken into account, the temperature of fireballs
is close to that of static pair equilibrium plasmas
obtained by past studies.
It is valuable to seek a much faster solution ($U > 1$)
than AT07.
As is usually seen in solutions of the accretion disk,
by varying the boundary conditions
we may obtain another type of solution with high temperature and high outward advection.
However, it is difficult to make $U > 1$ within the energy-release region,
because the effect of Compton drag may be crucial in such a compact region.
The photon field is produced from the plasma itself so that
the photon distribution is not sufficiently beamed in the inner region,
where the plasma is produced.
We expect plasma acceleration only in the outer region,
where the photon distribution is highly beamed.

Another possibility is nonsteady solutions of outflows.
We should note that the timescale of radiative transfer is 
much longer than the 
dynamical timescale.
The lifetime of accretion plasma may 
not be long enough to achieve the steady-state radiation field
obtained here and in AT07. 
As in the original idea of the Wien fireball model, 
an instantaneous heating may produce
a runaway production of relativistic pairs
because of slow annihilation in a hot plasma.
Therefore, time-dependent simulations should be
examined in a future work.

\begin{acknowledgments}
We appreciate the anonymous referee for his/her helpful advice.
This work is supported in part by Scientific Research Grants
(F.T. 18540239 and 20540231) from the Ministry of Education, Culture, 
Sports, Science and Technology of Japan.
\end{acknowledgments}


\end{document}